\documentclass[aps,prl,twocolumn,superscriptaddress,floatfix]{revtex4}
\usepackage{graphicx}
\usepackage{afterpage}
\usepackage{dcolumn}
\usepackage{bm}
\newcommand{\glue}{$\tilde{\Pi}(\omega)$ }

\newcommand{\hg}{HgBa$_{2}$CuO$_{4+\delta}$ }
\newcommand{\bitri}{Bi$_{2}$Sr$_{2}$Ca$_{2}$Cu$_{3}$O$_{10+\delta}$ }
\newcommand{\biduo}{Bi$_{2}$Sr$_{2}$CaCu$_{2}$O$_{8+\delta}$ }
\newcommand{\bimono}{Bi$_{2}$Sr$_{2}$Cu$_{2}$O$_{6+\delta}$ }

\begin{document}
\preprint{APS/123-QED}
\title{Observation of a robust peak in the glue function of the high-T$_c$ cuprates in the 50-60 meV range.}
\author{E. van Heumen}
\author{E. Muhlethaler}
\author{A.B. Kuzmenko}
\author{D. van der Marel}
\affiliation{D\'epartement de Physique de la Mati\`ere
Condens\'ee, Universit\'e de Gen\`eve, quai Ernest-Ansermet 24,
CH1211 , Gen\`eve 4, Switzerland}
\author{H. Eisaki}
\affiliation{Nanoelectronics Research Institute, National
Institute of Advanced Industrial Science and Technology,
Tsukuba, Japan}
\author{M. Greven}
\author{W. Meevasana}
\author{Z.X. Shen}
\affiliation{Department of Physics, Applied Physics, and
Stanford Synchrotron Radiation Laboratory, Stanford University,
Stanford, CA 94305}
\begin{abstract}
We take advantage of the
connection between the free carrier optical conductivity and
the glue function in the normal state, to reconstruct from the infrared optical
conductivity the glue-spectrum of ten different high-T$_c$
cuprates revealing a robust peak in the 50-60 meV range and a broad continuum
at higher energies for all measured charge carrier concentrations and temperatures
up to 290 K. We observe an intriguing correlation between the doping trend of the experimental glue spectra and the critical temperature.
\end{abstract}
\maketitle
%
%
%
%
The theoretical approaches to the high T$_c$ pairing mechanism in the cuprates are divided in two main groups: According to the first school electrons form pairs due to a retarded attractive interaction mediated by virtual bosonic excitations
in the solid\cite{scalapino-PRB-1986,varma-PRL-1989,millis-PRB-1990,dolgov-physc-1991,abanov-spec-2001}. These bosons can be lattice vibrations,
fluctuations of spin-polarization, electric polarization or
charge density.  The second school concentrates on a pairing-mechanism entirely due to the non-retarded Coulomb interaction\cite{anderson-sci-2007} or so-called Mottness\cite{phillips-annphys-2006}. Indeed, optical experiments have found indications for mixing of high and low energy degrees of freedom when the sample enters into the superconducting state\cite{basov-sci-1999,molegraaf-science-2002,carbone-PRB-2006a,carbone-PRB-2006b}.
\\
An indication that both mechanisms are present was obtained by Maier, Poilblanc and Scalapino\cite{maier-PRL-2008}, who showed that the 'anomalous' selfenergy associated with the pairing has a small but finite contribution extending to an energy as high as $U$, demonstrating that the pairing-interaction is, in part, non-retarded. They arrived at the conclusion, that "for the cuprate materials, the relative weight of the retarded and nonretarded interaction remains an open question. Thus, the continuing experimental search for a pairing glue in the cuprates is important and will play an essential role in determining the origin of the high-T$_c$ pairing interaction." Aforementioned glue can be expressed with the help of the density of states of these bosons multiplied by the electron-boson
coupling, $\alpha^2F(\omega)$ for phonons and $I^2\chi(\omega)$
for spinfluctuations, in this Letter represented by the general symbol
\glue. An important consequence of the electron-boson coupling is, that the energy of the quasi-particles
relative to the Fermi level, $\xi$, is renormalized, and
their lifetime becomes limited by inelastic decay processes
involving the emission of bosons. The corresponding energy
shift and the inverse lifetime, {\em i.e.} the real and
imaginary parts of the self-energy, are expressed as the
convolution of the 'glue-function' $\tilde{\Pi}(\omega)$ with a kernel
$K(\xi,\omega,T)$ describing the thermal excitations of the glue and the electrons\cite{kernel}
\begin{equation}
\Sigma(\xi)=\int K(\xi,\omega,T) \tilde{\Pi}(\omega)d\omega
  \label{sigma}
\end{equation}

In the absence of a glue and of scattering off impurities the
effect of applying an AC electric field to the electron gas is
to induce a purely reactive current response, characterized by
the imaginary optical conductivity
$4\pi\sigma(\omega)=i\omega_p^2/\omega$, where the plasma frequency, $\omega_p$,
is given by the (partial) f-sum rule for the conduction electrons. The effect of coupling
the electrons to bosonic excitations is revealed by a finite,
frequency dependent dissipation, which can be understood as
arising from processes whereby a photon is absorbed by the
simultaneous creation of an electron-hole pair and a boson. As
a result, the expression for the optical conductivity in the normal state,
$4\pi\sigma(\omega)=i\omega_p^2/\{\omega+2\Sigma_{opt}(\omega)\}$,
now contains a memory function\cite{gotze-PRB-1972} equivalent to an 'optical self-energy'. A particularly useful
aspect of this representation is that $\Sigma_{opt}(\omega)$
follows in a straightforward way from the experimental optical
conductivity. The optical
self-energy is related to the single particle self-energies by
the expression\cite{pballen-PRB-1971}
\begin{equation}
\frac{2\Sigma_{opt}(\omega)}{\omega}=
\left\{
 {\int \frac{f(\xi)-f(\xi+\omega)}{\omega+\Sigma^*(\xi)-\Sigma(\xi+\omega)}d\xi}\right\}^{-1}
 - 1
  \label{Kubo}
\end{equation}

The central assumption in the above is the validity of the Landau Fermi-liquid
picture for the normal state. The aforementioned strong coupling analysis is
therefore expected to work best on the overdoped side of the
cuprate phase diagram, where the state of matter appears to
become increasingly Fermi liquid like. If antiferromagnetism is necessary to
obtain the insulating state in the undoped parent compounds, as has
been argued based on the doping trends of the Drude spectral
weight\cite{millis-nphys-2008}, the strong coupling analysis
may in principle be relevant for the entire doping range studied. However,
in the limit of strong interactions aforementioned formalism
needs to be extended, {\em e.g.} with vertex corrections, and
eventually breaks down. We therefore {\em define} the function
\glue as the {\em effective} quantity which, in combination with Eqs. \ref{sigma} and
\ref{Kubo}, returns the exact value of $\Sigma_{opt}(\omega)$ for each frequency. Defined in this way \glue captures {\em all} correlation
effects regardless whether the system is a Fermi-liquid or not.
This becomes increasingly
relevant when the doping is lowered below optimal doping.

\begin{figure}[htb]
\centering
\includegraphics[width=8.5 cm]{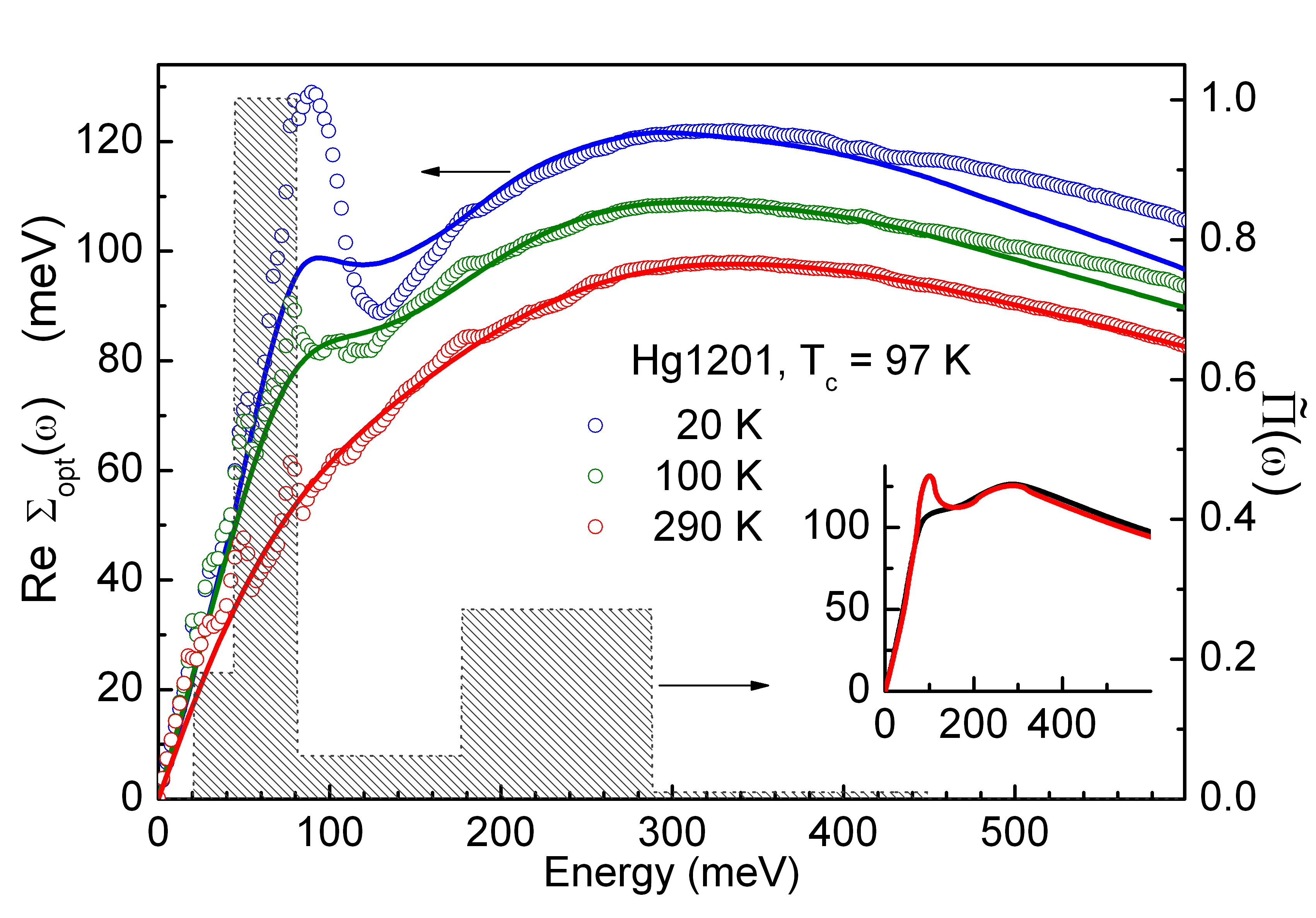}
\caption{Experimental optical self energy of \hg for 3 selected temperatures
(open circles). The solid curve at 290 Kelvin is obtained from a fit of \glue, shown as the dashed surface. The solid curves at 100 K and 20 K were calculated with the same \glue function corresponding to 290 Kelvin. This proves that the self energy feature between 80 and 100 meV (a shoulder at 100 K and a peak at 20 K) is caused by the prominent peak in \glue at approximately 60 meV. The sharpening of this feature at low temperature is due to the superconducting gap, an aspect not captured by Eq. \ref{Kubo} and therefore not reproduced in the calculated solid curves. In the inset the gap-induced sharpening is illustrated by the optical self energy without (black) and with (red) a 15 meV superconducting gap, calculated using Allen's relation\cite{pballen-PRB-1971}.}
\label{fig_selfenergy}
\end{figure}

Here we take advantage of the connection between the
temperature and frequency dependent conductivity in the normal
state and the glue-spectrum to test experimentally the
consequences of the standard approach, to check the internal
consistency of it, and to determine the range of doping where
internal consistency is obtained. For a d-wave superconductor,
the momentum dependence is essential to understand the details
of the pairing.  This, of course, is difficult to handle for
optical spectroscopy which is inherently a momentum integrated
probe. Nevertheless, optical spectra provide the important
information on the energy scale of the bosons involved and on
the doping and temperature evolution. We use a standard least
squares routine to fit a histogram representation of \glue to
our experimental infrared spectra. The quantity \glue is shown
in Fig. \ref{fig_selfenergy} for optimally doped \hg
(Hg-1201)\cite{heumen-PRB-2007} for $T= 290$ K, together with
the optical self energies calculated from this function at
three different temperatures. For 290 K the theoretical curve
runs through the data points, reflecting the full convergence
of the numerical fitting routine. It is interesting to notice,
that the shoulder in the experimental data at 100 K is
reproduced by the same \glue function as the one used to fit
the 290 K data. In other words, the strong temperature
dependence of the experimental optical spectra is entirely due
to the Fermi and Bose factors of Eqs. \ref{sigma} and
\ref{Kubo}. It can be excluded that the shoulder at 80 meV is
due to the pseudo-gap, since a gap is certainly absent for
temperatures as high as 290 K. The shoulder is therefore
entirely due to coupling of the electrons to a mode at
approximately 60 meV. On the other hand, the considerable
sharpening of this feature for temperatures lower than 100 K
finds a natural explanation in the opening of a gap, as
illustrated in the inset of Fig. \ref{fig_selfenergy}. This
example confirms the close correspondence between the features
in $\Sigma_{opt}(\omega)$ and in $\tilde{\Pi}(\omega)$ pointed
out in Ref. [\onlinecite{norman-PRB-2006}]. In particular the
broad maximum in $\Sigma_{opt}(\omega)$ has its counterpart in
the high intensity region of \glue terminating at 290 meV.

\begin{figure*}[t!]
\centering
\includegraphics[width=17 cm]{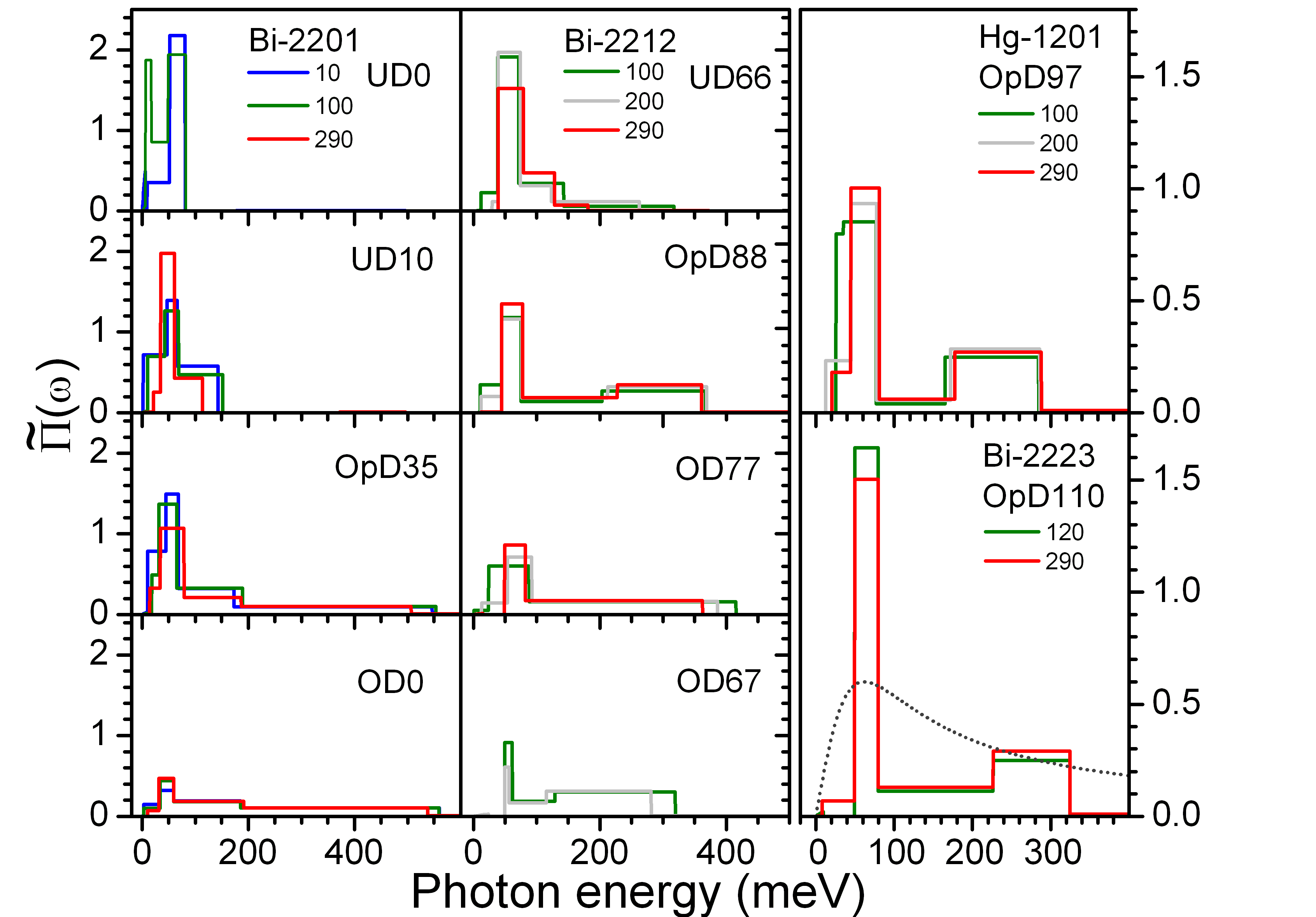}
\caption{Electron-boson coupling function \glue for Bi-2201 at
four different charge carrier concentrations (10 K, 100 K, 290 K),
Bi-2212 at four charge carrier concentrations, and optimally doped
Bi-2223 and Hg-1201 (100 K, 200 K, 290 K). The dotted curve in
the lower right panel represents the spin-fluctuation model\cite{norman-PRB-2006,millis-PRB-1990}.}
\label{fig_glue}
\end{figure*}

As summarized in Fig. \ref{fig_glue}, we have analyzed previously published optical spectra of 6 different
samples belonging to different families of materials, {\em i.e.} optimally doped Hg-1201\cite{heumen-PRB-2007} and \bitri
(Bi-2223)\cite{carbone-PRB-2006b}, as well as four  \biduo (Bi-2212)
crystals \cite{molegraaf-science-2002,carbone-PRB-2006a} with different hole concentrations. In addition, we analyzed new data for four \bimono (Bi-2201) crystals with different hole
concentrations\cite{heumen-PRB-inprep}.

Excellent fits were
obtained for all temperatures, but the \glue spectra exhibit a significant temperature dependence,
in particular at the low frequency side of the \glue spectrum. Since all thermal factors
contained in Eqs. \ref{sigma} and \ref{Kubo} are, in principle,
folded out by our procedure, the remaining temperature
dependence of \glue reflects the thermal properties of the
'glue-function' itself. Such temperature dependence is a direct consequence of the peculiar DC and far infrared conductivity, in particular the $T$-linear DC resistivity and $\omega/T$ scaling of $T\sigma(\omega,T)$ at optimal doping\cite{dirk-nature-2003}. For the
highest doping levels both \glue and its temperature dependence
diminish, which is an indication that a Fermi liquid
regime is approached. The most strongly underdoped sample, Bi-2201-UD0, exhibits an upturn of the imaginary part of the experimental optical self-energy for $\omega\rightarrow 0$. This aspect of the data can not be reproduced by the strong coupling expression, resulting in an artificial and unphysical peak at $\omega\approx 0$ of the fitted \glue function.

\begin{table}[h]
\begin{center}
\begin{tabular}{l|l|llll|llll|l|l}
\hline
x&&0.09&0.11&0.16&0.22&0.11&0.16&0.20&0.21&0.16&0.16\\
\hline
$T_c$&K&0&10&35&0&66&88&77&67&110&97\\
\hline
$\hbar\omega_p$ & eV         &1.75&1.77&1.92&1.93&2.36&2.35&2.45&2.33&2.43&2.10\\
$\hbar\tilde{\omega}$ &meV   & -  &70&81&103&92&124&116&154&101&81\\
$\lambda$         &      & -  &2.96&2.96&1.41&2.65&2.14&1.40&0.97&2.18&1.85\\
\hline
\end{tabular}
\end{center}
\caption{\label{sampleparameters} Strong coupling parameters of the ten compounds. The hole-doping is indicated on the first line. From left to right: Bi2201 (columns 1 to 4), Bi2212 (columns 5 to 8), Bi2223 (8th column) and Hg1201 (column 10). The definition of $\tilde{\omega}$ is
$\ln(\tilde{\omega})=2\lambda^{-1}\int_0^{\infty} \omega^{-1}\tilde{\Pi}(\omega)\ln(\omega)d\omega$. }
\label{table1}
\end{table}

We observe two main features in the glue-function: A robust
peak at 50-60 meV and a broad continuum. The upper limit of \glue is situated around
approximately 300 meV for optimally doped single layer Hg1201, and for the bilayer and trilayer samples. Interestingly the continuum extends to the highest energies (550 meV for the single-layer samples and 400 meV for the bilayer) for the weakly overdoped samples, whereas the continuum of the strongly doped bilayer sample extends to only 300 meV. There is also a clear trend of a contraction of the continuum to lower energies when the carrier concentration is reduced. Taken together this doping trend suggests a possible relation of the glue-function to a quantum critical point at a doping concentration slightly higher than optimal doping\cite{tallon-proc-1999,dirk-nature-2003}. Hence, part of the glue function has an energy well above the upper
limit of the phonon frequencies in the cuprates ($\sim$ 100 meV). Consequently
the high energy part of \glue reflects in one way or another
the strong coupling between the electrons themselves. The coupling constant
is obtained from the relation $\lambda=2\int_{0}^{\infty}\tilde{\Pi}(\omega)/\omega d\omega$.
The result (see Table \ref{table1}) shows a strong and systematic increase of
$\lambda$ for decreasing hole concentration, which probably requires a
theoretical treatment beyond the strong coupling expansion outlined in the introduction.

The most prominent feature, present in all spectra reproduced
in Fig. \ref{fig_glue}, is a peak corresponding to an average
frequency of 55 meV. Perhaps the most striking aspect of this
peak is the fact that its energy is practically independent of
temperature (up to room temperature) and sample composition.
Moreover, the intensity and width are essentially temperature independent. While our
results confirm by and large the observations of Hwang
\textit{et al.} in the pseudo-gap phase\cite{timusk4,timusk5}, the
persistence of the 50-60 meV peak to room temperature has not been
reported before. This peak arises most likely from the same boson that is
responsible for the 'kink' seen in angle
resolved photoemission (ARPES) experiments along the nodal
direction in k-space at approximately the same
energy\cite{bogdanov-PRL-2000,lanzara-NAT-2001,non-PRL-2006}.
The peak-dip-hump structure in the tunneling spectra
(STS)\cite{lee-nat-2006,levy-condmat-2007,zasad-PRL-2001} has
also been reported at approximately the same energy.

\begin{figure}[t!]
\centering
\includegraphics[width=8.5 cm]{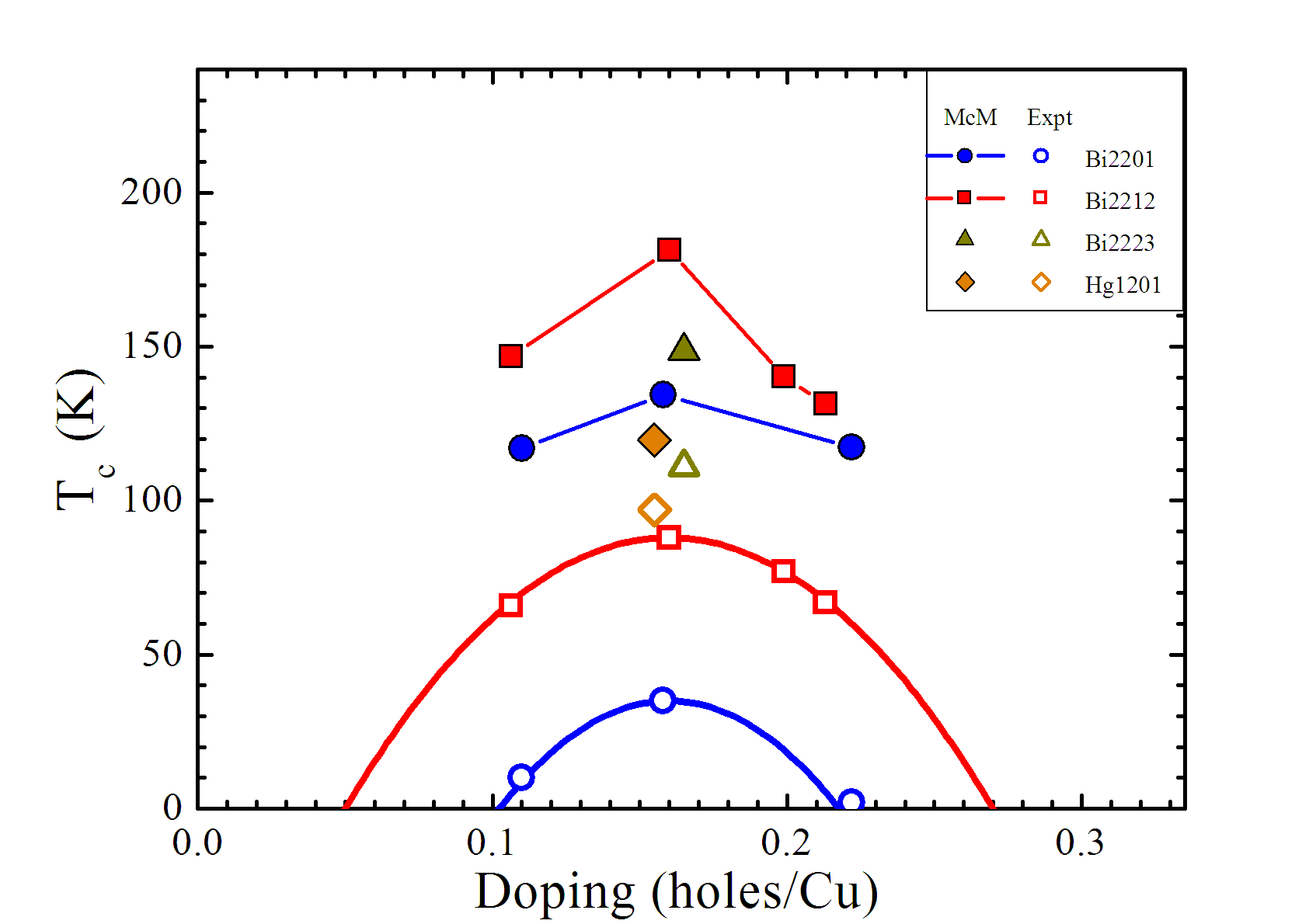}
\caption{Experimental critical temperature and $T_c$'s
calculated from the McMillan formula using the experimentally
measured \glue of Fig. \ref{fig_glue} at 290 Kelvin as input
parameters. }
\label{fig_tc}
\end{figure}

It is interesting to correlate our experimentally obtained glue-function with $T_c$. Millis, Varma and Sachdev\cite{millis-PRB-1988} have shown, that, if \glue can be decomposed in an s-wave and a d-wave channel and $\mu^*$ is considered negligibly small for the d-wave
channel, then the critical temperature of the d-wave
superconducting phase transition is $T_c \le 0.83 \tilde{\omega}
\exp(-(1+\lambda)/\lambda)$ where
$\tilde{\omega}$ is the logarithmic average of the frequency in
\glue. We calculated this upper limit using the experimental values indicated in Table \ref{table1}. The $T_c$'s
with this formula are in the 100-200 K range for all
samples studied here, and, as shown in Fig. \ref{fig_tc},
they correlate with the experimentally observed doping trends of
$T_c$ for the single-layer and bilayer high-T$_c$ materials.
The calculated critical temperatures are about twice as large
as the experimentally observed ones for the Bi2212 series,
whereas in the Bi2201 single layer compounds $T_c$ appears to
be strongly suppressed as compared to the calculated value. The low value of T$_c$ in single layer Bi2201 correlates with a relatively high intensity of \glue below the 50-60 meV peak, and a very low intensity or absence of the 0.25 eV peak.
In contrast, the three samples with the highest critical temperature, Bi2212 OpD88, Bi-2223 and Hg-1201, show the most pronounced weight at $\sim$ 0.25 eV.

In summary, the \glue spectrum obtained from the optical spectra of 10 different compounds using a strong coupling analysis, is observed
to consist of two features: (i) a
robust peak in the range of 50 to 60 meV and (ii)
a doping dependent continuum extending to 0.3
eV for the samples with the highest $T_c$. Finally, we observe an intriguing correlation between the doping trend of the experimental glue spectra and the critical temperature.

We gratefully acknowledge C.M. Varma, D.J. Scalapino, J.
Zaanen, A.V. Chubukov, C. Berthod, J.C. Davis, and A. Millis for
stimulating discussions. This work is supported by the
Swiss National Science Foundation through Grant No.
200020-113293 and the National Center of Competence in
Research (NCCR) "Materials with Novel Electronic
Properties—MaNEP".

\section{Appendix}
The inversion of Eq's 1 and 2 allows to extract \glue from
experimental data of the optical conductivity, or related
optical spectra. The accuracy of the resulting \glue spectrum
is in practice limited by the convolution with thermal factors
expressed by Eq's 1 and 2 \cite{dordevic-PRB-2005}. Microscopic
models giving roughly the same \glue spectra, which differ
however in the details of the frequency dependence of this
quantity, may therefor provide fits to the directly measured
optical quantities, such as infrared reflectance spectra, which
at first glance look satisfactory, but the remaining
discrepancies with the experimental spectra may nevertheless be
of significant importance for the proper understanding of the
optical data. It is therefore of crucial importance to test the
'robustness' of each fit with regard to the spectral shape of
the \glue function imposed by such models. This robustness can
be tested by including in the fit-routine one or several
'oscillators' superimposed on the model function. When the
model glue function provides a complete description of the
electronic structure, adding extra oscillators will not result
in an improvement of the quality of the fit. We have used this
approach to test functional forms commonly used in the
literature, in particular the marginal Fermi liquid (MFL) model
\cite{varma-PRL-1989} and the Millis-Monien-Pines (MMP)
representation of the spin fluctuation
spectrum\cite{millis-PRB-1990}. We found that neither of these
functional forms describe completely the experimental data.

In search of a more flexible form of \glue we used a
superposition of lorentzian oscillators and found that it could
be used to describe all available experimental data in a
consistent manner. The resulting \glue functions and trends are
equivalent to those in Fig. \ref{fig_glue}. From these initial
tests we concluded that due to the thermal smearing expressed
by Eq's 1 and 2 our \glue spectra can only be determined with
limited resolution. This lead us to the use of a histogram
representation, where each block in the histogram represents a
likelihood to find coupling to a mode with a well determined
coupling strength. Only for the lowest frequency interval ($0 <
\omega < \omega_1$) a triangular shape was used instead of a
block, which is necessary to avoid problems with the
convergence of the integral $\lambda=2\int_0^{\infty}
\omega^{-1}\tilde\Pi(\omega) d\omega$. In practice the output
generated by the fitting routine has low intensity in this
first interval, and the triangles are therefore difficult to
distinguish in Fig. \ref{fig_glue}.

To give an example: The block centered at 55 meV seen in the
Hg-1201 sample in Fig. \ref{fig_glue} has $\lambda \sim 1$ and
a width of about 30 meV. Our histogram representation implies
the presence of a coupling to one or several modes between 45
meV and 75 meV with an integrated coupling strength of 1. The
histograms thus constitute the most detailed representation of
\glue given the precision of our experimental reflectivity and
ellipsometry spectra.

\begin{figure}[h]
\centering
\includegraphics[width=8.5 cm]{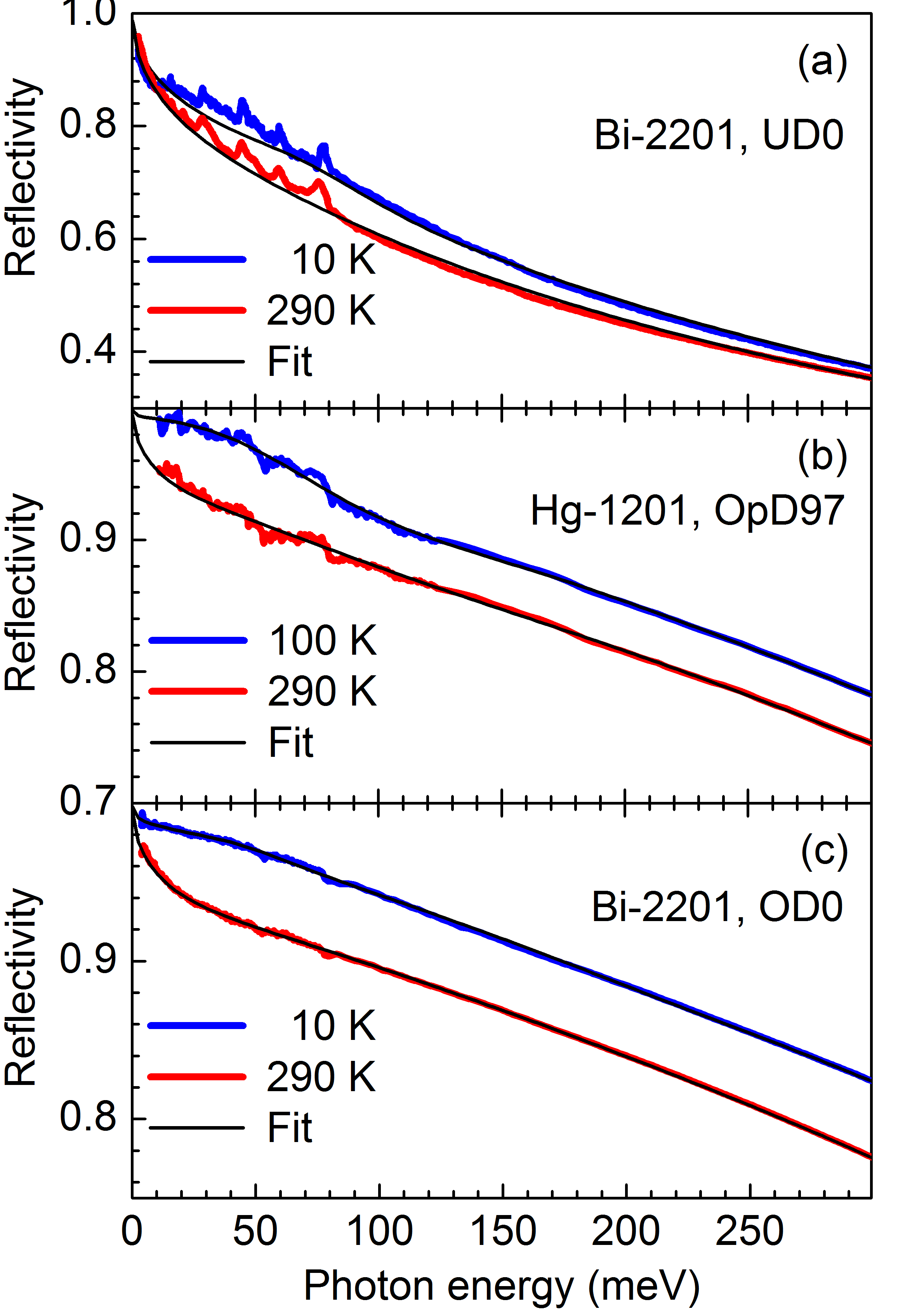}
\caption{Experimental reflectivity (red and blue lines) and fit curves (black lines) for selected samples and temperatures.
(a): underdoped non-superconducting Bi-2201 (UD0). (b): optimally doped Hg-1201 with $T_{c}\approx$ 97 K (OpD97) [\onlinecite{heumen-PRB-2007}]
(c): overdoped non-superconducting Bi-2201 (OD0). Weak sharp peaks, particularly visible for the strongly underdoped sample in panel (a) are due to transverse optical phonons, which we do not intend to fit.}
\label{fig_fitquality}
\end{figure}

Examples of experimental reflectivity data together with the
fits are shown in Fig. \ref{fig_fitquality} for a selection of
representative data sets spanning the entire doping and
temperature range. As the fitted curves are within the limits
of the experimental noise, further reduction of $\chi^2$, while
in principle possible by fitting the statistical noise of the
data, can not improve the accuracy of the \glue functions.
Starting from a \glue  function we can calculate the optical
conductivity, which in turn is fed into standard Fresnel
expressions to calculate the experimentally measured
quantities, {\em i.e.} reflectivity and ellipsometric
parameters. The fitting routine is based on the
Levenberg-Marquardt algorithm and uses analytical expressions
for the partial derivatives of the reflectivity coefficient
$R$, and the ellipsometric parameters $\psi$ and $\Delta$
relative to the parameters describing the \glue function.
\begin{figure*}[t] \centering
\includegraphics[width=17 cm]{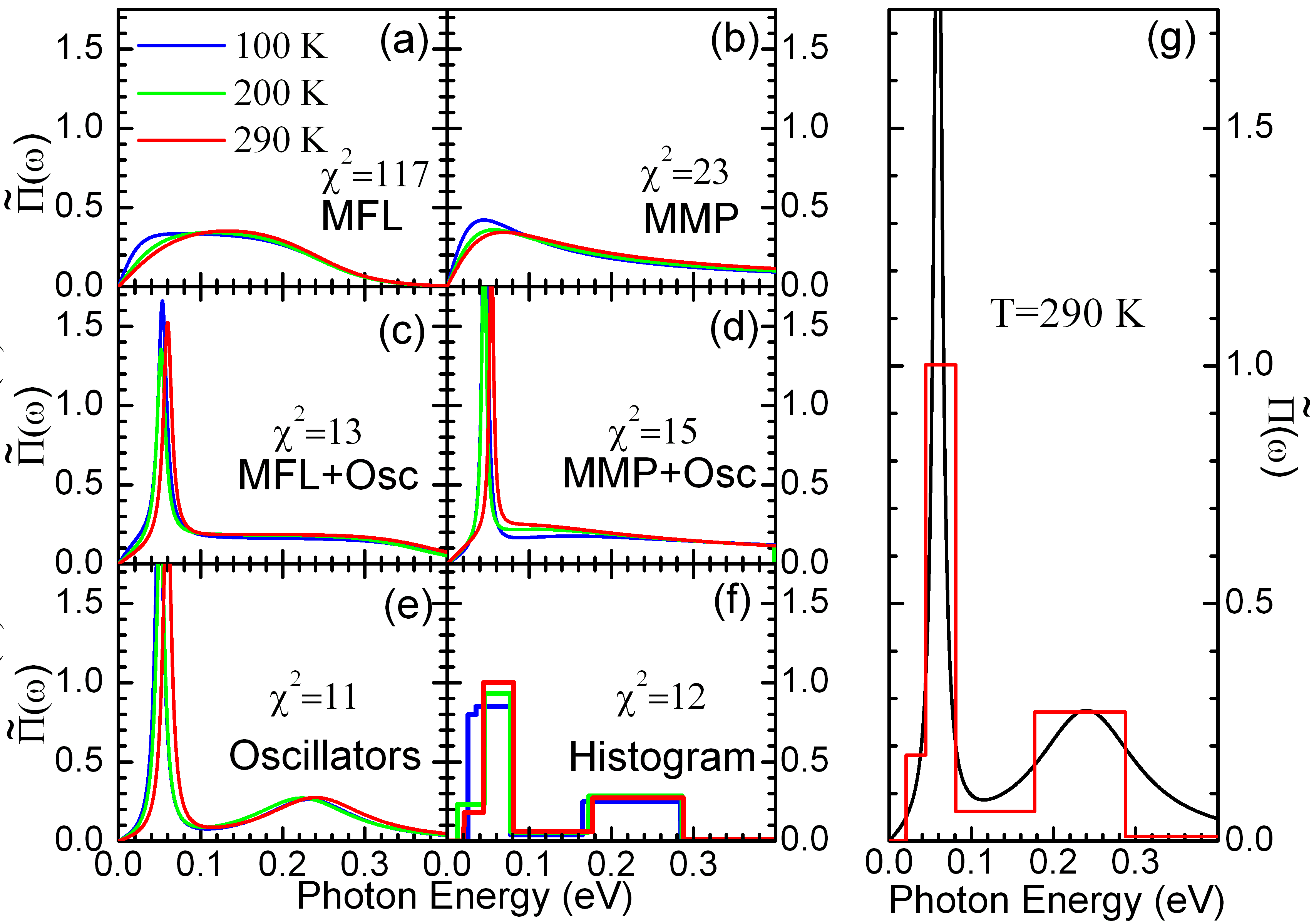}
\caption{Comparison of several models. The number of data
points over which $\chi_j^2$ is summed (see Eq. 1) is $N=400$.
The quoted values of $\chi^{2}$ are those for the room
temperature spectra. At 100 K these values increase by a factor
1.5 (MFL) and 3.5 (MMP). In contrast, in the oscillator and the
histogram model the $\chi^{2}$ was found to be independent of
temperature.} \label{fig_models}
\end{figure*}
The algorithm is based around minimizing a functional
$\chi^{2}$ which is given by,
\begin{equation}
\chi^{2}=\sum_{i=1}^{N}\left(\frac{R(\omega_{i})-f(\omega_{i},p_{1},...,p_{n})}{\sigma_{i}}\right)^{2}
\end{equation}
where $R(\omega_{i})$ is an experimentally measured datapoint,
$f(\omega_{i},p_{1},...,p_{n})$ is the calculated value in this
point based on parameters $p_{1},...,p_{n}$ and the difference
between these two is weighed by the errorbar $\sigma_{i}$
determined for $R(\omega_{i})$. For a given set of reflectivity
and ellipsometry data at one particular temperature, using a
standard PC, the iteration takes about 3 hours until
convergence is reached. Although the Levenberg-Marquardt least
squares method is an extremely powerful method to find the
minimum of $\chi^{2}$ in a multidimensional parameter space it
also has difficulties with finding the best solution to the
non-linear problem Eq's 1 and 2. For each individual sample and
temperature displayed in Fig. \ref{fig_models} several tests
have been performed where in each test the optimization process
was started from a different set of starting parameters to
ensure that $\chi^{2}$ has converged to the global minimum in
parameter space. To give some idea of the robustness of our
method we will here discuss one representative example:
optimally doped Hg-1201.

The models are evaluated based on the minimum found for
$\chi^{2}$. A comparison of Fig. \ref{fig_models} (a-d) shows
that the MMP model describes better the optical data then the
MFL model but that they give similar results if we add an extra
oscillator to these models. Panels S\ref{fig_models}e-f show
the model independent results mentioned above and are very
similar to the modified MMP and MFL model. The models in these
last two panels have the same $\chi^{2}$ and the comparison in
Fig. \ref{fig_models}g shows that the histogram representation
realistically expresses the uncertainty in the position of the
low energy peak, while the correspondence between the features
in both models remains excellent. It is interesting that the
model with two oscillators is described by 6 parameters, while
the histogram representation uses 12 parameters. The fact that
the fit-routine adjusts the latter 12 parameters in such a
manner as to produce in essence the two oscillator lineshape,
proves that the features represented in the righthand panel of
Fig. \ref{fig_models} are realistic. An extensive discussion of
this analysis is provided in Ref. \cite{heumen-LT-2009}.


\begin{thebibliography}{99}

\bibitem{scalapino-PRB-1986}
    D.J. Scalapino, E. Loh, J.E. Hirsch,
    \textit{Phys. Rev. B} \textbf{34,} 8190 (1986).

\bibitem{varma-PRL-1989}
    C.M. Varma,
    {\em et al.}, \textit{Phys. Rev. Lett.} \textbf{63,} 1996 (1989).

\bibitem{millis-PRB-1990}
    A.J. Millis, H. Monien, D. Pines,
    \textit{Phys. Rev. B} \textbf{42,} 167 (1990).

\bibitem{dolgov-physc-1991}
    S.V. Shulga, O.V. Dolgov and E.G. Maksimov,  \textit{Physica} C
    \textbf{178}, 266 (1991).

\bibitem{abanov-spec-2001}
    Ar. Abanov, A.V. Chubukov, J. Schmalian,
    \textit{J. Elec. Spec. Rel. Phen.} \textbf{117,} p129
    (2000).

\bibitem{anderson-sci-2007}
    P. W. Anderson, Science {\bf 316}, 1705 (2007).

\bibitem{phillips-annphys-2006}
    P. Phillips
    \textit{Ann. Phys.} \textbf{321,} 1634 (2006).

\bibitem{basov-sci-1999}	
    D. N. Basov  et al., Science 283, 49-51 (1999).

\bibitem{molegraaf-science-2002}
     H.J.A. Molegraaf, C. Presura, D. van der Marel, P.H. Kes, M. Li,
    \textit{Science} \textbf{295,} 2239 (2002).

\bibitem{carbone-PRB-2006a}
    F. Carbone \textit{et al.},
    \textit{Phys. Rev. B} \textbf{74,} 064510 (2006).

\bibitem{carbone-PRB-2006b}
    F. Carbone \textit{et al.},
    \textit{Phys. Rev. B} \textbf{74,} 024502 (2006).

\bibitem{maier-PRL-2008}
    T. A. Maier, D. Poilblanc, and D. J. Scalapino,
    \textit{Phys. Rev. Lett.} \textbf{100,} 237001 (2008).

\bibitem{kernel} $K(\xi,\omega,T)= \int\left[\
    \frac{n(\omega)+f(\epsilon)}{\xi-\epsilon+\omega+i\delta} +
    \frac{n(\omega)+1-f(\epsilon)}{\xi-\epsilon-\omega-i\delta}
    \right]d\epsilon $
    where $n(\omega)$ and $f(\epsilon)$ are the Bose and Fermi-Dirac distribution functions respectively\cite{pballen-PRB-1971}.

\bibitem{gotze-PRB-1972}
    W. Goetze, P. Woelfle,
    \textit{Phys. Rev. B} \textbf{6,} 1226 (1972).

\bibitem{pballen-PRB-1971}
    P.B. Allen,
    \textit{Phys. Rev. B} \textbf{3,} 305, (1971).

\bibitem{millis-nphys-2008}
    A. Comanac, L. de' Medici, M. Capone, A.J. Millis,
    \textit{Nature Phys.} \textbf{4,} 287, (2008).

\bibitem{heumen-PRB-2007}
    E. van Heumen \textit{et al.},
    \textit{Phys. Rev. B} \textbf{75,} 054522 (2007).

\bibitem{norman-PRB-2006}
    M.R. Norman, A.V. Chubukov,
    \textit{Phys. Rev. B} \textbf{73,} 140501 (2006).

\bibitem{heumen-PRB-inprep}
    E. van Heumen {\em et al.}, in preparation.

\bibitem{dirk-nature-2003}
    D. van der Marel \textit{et al.}
    \textit{Nature} \textbf{425,} 271 (2003).

\bibitem{tallon-proc-1999}
    J.L. Tallon \textit{et al.}
    \textit{phys. stat. sol. (b)} \textbf{215,} 531 (1999).

\bibitem{timusk4} 
    J. Hwang, T. Timusk, E. Schachinger, J.P. Carbotte,
    \textit{Phys. Rev. B} \textbf{75,} 144508 (2007).

\bibitem{timusk5} 
    J. Hwang, E.J. Nicol, T. Timusk, A. Knigavko, J.P. Carbotte,
    \textit{Phys. Rev. Lett.} \textbf{98,} 207002 (2007).

\bibitem{bogdanov-PRL-2000}
     P.V. Bogdanov \textit{et al.},
     \textit{Phys. Rev. Lett.} \textbf{85,} 2581 (2000).

\bibitem{lanzara-NAT-2001}
    A. Lanzara \textit{et al.},
    \textit{Nature} \textbf{412,} 510 (2001).

\bibitem{non-PRL-2006}
    W. Meevasana \textit{et al.},
    \textit{Phys. Rev. Lett.} \textbf{96,} 157003 (2006).

\bibitem{lee-nat-2006}
    J. Lee \textit{et al.},
    \textit{Nature} \textbf{442,} 546 (2006).

\bibitem{levy-condmat-2007}
    G. Levy de Castro {\em et al.}, cond-mat/0703131.

\bibitem{zasad-PRL-2001}
    J.F. Zasadzinski \textit{et al.},
    \textit{Phys. Rev. Lett.} \textbf{87,} 067005 (2001).

\bibitem{millis-PRB-1988}
    A.J. Millis, C.M. Varma, S. Sachdev
    \textit{Phys. Rev. B} \textbf{37,} 4975 (1988).

\bibitem{dordevic-PRB-2005}
    S.V. Dordevic \textit{et al.},
    \textit{Phys. Rev. B} \textbf{71,} 104529 (2005).

\bibitem{heumen-LT-2009}
    E. van Heumen, A.B. Kuzmenko, D. van der Marel,
    submitted to the proceedings of LT25, to be published in \textit{J. Phys.: Conf. Ser.}.

\end{thebibliography}
\end{document}